\documentstyle[11pt,aaspp4]{article}

\begin{document}
\title{NICMOS Imaging of Molecular Hydrogen Emission in 
Seyfert Galaxies}

\author{
A.~C.\ Quillen\altaffilmark{1}$^,$\altaffilmark{2}, \&
A.~Alonso-Herrero\altaffilmark{1},
M.~J.~Rieke\altaffilmark{1},
G.~H.~Rieke\altaffilmark{1},
M.~Ruiz\altaffilmark{3},  \&
V.~Kulkarni\altaffilmark{1}
}
\altaffiltext{1}{The University of Arizona, Steward Observatory, Tucson, AZ 85721}
\altaffiltext{2}{E-mail: aquillen@as.arizona.edu}
\altaffiltext{3}{Department of Physical Sciences, University of Hertfordshire, 
College Lane, Hatfield, Herts AL10 9AB, UK}

\begin{abstract}
We present NICMOS imaging of broad band and molecular hydrogen
emission in Seyfert galaxies. In  6/10 Seyferts we detect
resolved or extended emission in the 
1-0 S(1)$\lambda 2.121\micron$ or 1-0 S(3)$\lambda 1.9570\micron$ molecular
hydrogen lines.  We did not detect emission in the most distant galaxy
or in the 2 Seyfert 1 galaxies in our sample because
of the luminosity of the nuclear point sources.
In NGC~5643, NGC~2110 and MKN~1066, molecular hydrogen emission
is detected in the extended narrow line region on scales
of a few hundred pc from the nucleus.  Emission is 
coincident with [OIII] and H$\alpha$+[NII] line
emission.  This emission is also
near dust lanes observed in the visible to near-infrared 
color maps suggesting that a multiphase medium exists near
the ionization cones and that the morphology of the line emission is
dependent on the density of the ambient media.
The high 1-0~S(1) or S(3) H$_2$ to H$\alpha$ 
flux ratio suggests that shock excitation of molecular hydrogen 
(rather than UV fluorescence) is 
the dominant excitation process in these extended features.
In NGC~2992 and NGC~3227 the molecular hydrogen emission
is from 800 and 100 pc diameter `disks' (respectively) which
are not directly associated with [OIII] emission and are 
near high levels of extinction ($A_V \gtrsim 10$).
In NGC~4945 the molecular hydrogen emission appears to be from 
the edge of a 100 pc superbubble.
In these 3 galaxies the molecular gas 
could be excited by processes associated with local star formation.
We confirm previous spectroscopic studies
finding that no single mechanism is likely to be responsible
for the molecular hydrogen excitation in Seyfert galaxies.
\end{abstract}


\section {Introduction}

The emission from molecular hydrogen (H$_2$) 
can be used to probe the distribution of excited dense molecular
gas in AGNs (active galactic nuclei).
H$_2$ emission (e.g., the 1-0~S(1) line at 2.121$\micron$)
is known to be bright in Seyferts (e.g., \cite{veilleux}; \cite{ruiz};
\cite{mouri}; \cite{koornneef}).  
Ultraviolet (UV) fluorescence (e.g., \cite{vandish}), 
excitation by low velocity shocks (e.g, \cite{draine}; \cite{hollenbach}),
and heating caused by X-rays 
are the primary emission mechanisms considered for the excitation
of molecular hydrogen (e.g., \cite{puxley}; \cite{veilleux}; \cite{mouri},
\cite{moorwood88},
and references theirin).  These mechanisms require a source of dense gas 
(density $\gtrsim 10^3$ cm$^{-3}$)
to be either located near a source of illumination
(an AGN or a starburst)
or actively affected by slow shocks induced either by jets or by
kinetic processes (winds, superwinds and supernovae) 
resulting from star formation.
Ground-based spectroscopic studies of Seyferts and LINERS have found that
the 1-0~S(1) flux weakly correlates with [FeII] 1.64 $\mu$m emission
(e.g., \cite{larkin}; \cite{ruiz}; \cite{mouri})
and 6cm radio flux density (\cite{forbes}; \cite{koornneef}; \cite{veilleux})
but with a substantial scatter.  No correlation between
the 1-0 S(1) line and hard X-ray flux in Seyferts 
has been observed (\cite{ruiz};
\cite{veilleux}) suggesting that heating by X-rays cannot
be the dominant cause of excitation.
These results suggest that more than one process causes
the H$_2$ emission.

High resolution imaging of line emission
makes it possible to discriminate between likely emission causes.  
For example, the morphological 
association between the narrow line region emission
lines (e.g., [OIII]$\lambda 5007$\AA) gas and the radio emission 
suggests a connection between radio ejecta and the ambient
gas (e.g., \cite{wilsonwillis}; \cite{unger}; \cite{pogge}).  
However, the difference in angle between jet axes, ionization 
axes and galaxy axes (\cite{nagar}) suggest that the distribution
of ambient gas in the galaxy also plays an important role in 
determining the morphology of the ionization cones
(also see \cite{quillen99} for examples of direct spatial associations).
The distribution of extended 
molecular emission therefore provides a clue to the source
of dense gas either illuminated by the central
source, affected by energetic motions caused by a jet,
or excited by processes associated with star formation
(e.g., \cite{maiolino}).


In this paper we present a high angular resolution imaging study of 
molecular hydrogen emission in Seyfert galaxies.
Images were observed with the NICMOS cameras on board the Hubble Space
Telescope (HST).  
We construct high angular resolution visible to infrared color maps from WFPC2
and NICMOS broad band images.
Red features caused by extinction in these maps are likely to 
trace dust lanes containing cold dense gas.  We compare
the morphology of the molecular hydrogen emission with that
seen in the color maps and when possible that of
[OIII] and H$\alpha$ + [NII] emission from ionized gas.

\section{Observations}


Our sample consists of 3 Seyfert 1 and 7 Seyfert 2
galaxies at a variety of inclinations.
We chose nearby Seyferts with bright 1-0~S(1) H$_2$ fluxes
from \cite{ruiz_}, \cite{veilleux_} and \cite{koornneef_}.
Redshifts were restricted so that either the 1-0~S(1) or 1-0~S(3) line
could be observed with existing NICMOS narrow band filters.
The H$_2$ 1-0~S(3) 1.9576 $\mu$m line is approximately as bright as the
1-0~S(1) line at 2.1213 $\mu$m in Seyfert galaxies (\cite{vandish}).
The sample is listed in Table 1.



Because the continuum in these galaxies is bright and contains structure
which can vary with position, 
we required observations with a narrow band filter near (in wavelength) to
the narrow band filter containing the emission line
to accurately subtract the continuum.
One HST orbit was spent per galaxy with approximately 
128s spent in the broad band filter (F205W or F222M) when possible 
and the remainder of the orbit evenly divided between
the two narrow band filters.

The galaxies were observed in
each filter at 4 different positions on the sky with a separation of
$0''.57$ or $1''.12$ apart for Camera 2 and 3 observations
respectively.  Images were reduced with the nicred data reduction
software (\cite{mcl}) using on orbit darks
and flats.  Each set of 4 images in a given filter
was then combined according to the position observed.
The pixel size for the NICMOS camera 2 is $\sim 0''.076$
and for the NICMOS camera 3 is $\sim 0''.204$.
Flux calibration for the NICMOS images
was performed using the conversion factors based on
measurements of the standard stars P330-E and P172-D during
the Servicing Mission Observatory Verification program 
and subsequent observations (M.~Rieke, private communication).

\subsection{Broad band and archival images}

For most of the galaxies we were able to obtain broad band near-IR
NICMOS images.  Three of the galaxies 
(MKN~6, NGC~5643, and MKN~1066) were part of snap shot 
observation programs
to observe at 1.60$\mu$m (in the F160W filter). 
NGC3227 and IC4329A are part of a GTO (guaranteed time observing) program. 
For comparison to the F160W broad band images
we were able to include as part of our program observations
at 2.05$\mu$m (in the F205W filter) of 
MKN~1066, NGC~4945, NGC~5506 and NGC~5643 and at 2.22$\mu$m 
(in the F222M filter) of MKN~938.

Broad band WFPC2 images in F814W (at $0.814\mu$m for NGC~5643)
and F606W ($0.606\mu$m for NGC~2110,  NGC~5643, NGC~2992, MKN~938, MKN~1066
and MKN~6) were retrieved from the HST archive and 
used with NICMOS images to create optical to infrared color maps.
When possible archive images were used to display
structure in the ionization cones in either 
H$\alpha$+[NII]$\lambda$6548, 6583\AA 
or [OIII]$\lambda$5007\AA. For the studies discussing these HST data 
see \cite{mulchaey94a_} on NGC~2110, \cite{simpson_} on NGC~5643,
\cite{bower95_} on MKN~1066, and \cite{capetti_} on MKN~6.  

\subsection{Narrow band images}

The molecular hydrogen emission was observed with the F212N,
F215N, F216N filters on Camera 2 or the F196N on Camera 3
depending on the observed (redshifted) wavelength of the object (see Table 1).
For some of the objects in our list, the 1-0~S(1) 2.1213$\micron$
line could not be
observed but the 1-0~S(3) $1.9570\micron$ line which is nearly as bright
as the 1-0~S(1) line could be observed with the F196N filter on Camera 3.
For all of the objects observed with the F196N filter
the line is on the red side of the filter
so there is no danger of confusion caused by the [SiVI] 1.96$\mu$m 
line which would
be outside the filter and should be narrow in the Seyfert 2s
(\cite{giannuzzo}).
NGC~3227 could be observed in either H$_2$ line but it
has broad lines. To reduce possible confusion from the [SiVI]
we chose to observe the 1-0~S(1) line.  For this object the line
lies at a position in the filter with only 60$\%$ of maximum transmission
but the object is bright enough so that good signal to noise
was achieved with the same integration time as the other objects.

To construct H$_2$ line emission maps, the narrow band continuum image 
(lacking the line) was scaled and subtracted from the narrow band image
containing the line.  From broad band images 
we estimated the spectral index (or color) of the continuum.  We then 
checked that variations in this index were unlikely to be responsible 
for the variations seen in the continuum subtracted image.
This was the case for all galaxies except for NGC~2992
and except for the nuclei where the structure of 
the point spread function makes it difficult
to measure intensities and colors. 
In NGC~2992 about 50\% of the variations seen 
in the continuum subtracted image 
could have been a result of large changes in the continuum spectral
index due to reddening.  We therefore corrected
the continuum image before subtracting it to account
for variations in this index.
The factor as a function of position used to correct each pixel 
in this image was derived from the spectral index 
which we measure from the ratio of the F606W and F205W images of this galaxy.
In no cases do the resulting line emission maps
show structure observed in the color maps.  
This suggests that they do
indeed trace the distribution of the 1-0 S(1) or S(3) lines.
To estimate the line fluxes we took into account the transmission
of the filter at the redshifted wavelength of the line.
Fluxes and surface brightnesses were measured in various
regions from the continuum subtracted images and are listed
in Table 2.  Fluxes agree within a factor of 2 of the published values
(listed in Table 1) in similar apertures.  
We measure higher nuclear fluxes for NGC~2992 and NGC~5643
than the listed values.  Some of the discrepancy
could be due to extreme nuclear continuum colors.

\subsubsection{NGC~3227}

NGC~3227 contains a bright nuclear point source which hampers
our ability to detect extended line emission.
We attempted to subtract a nuclear point source using
a model point spread function (PSF) created by Tiny Tim 
(\cite{tinytim}, \cite{krist}).
However the point source subtracted
narrow band images did not produce a line emission map 
significantly better than that using the images containing
the nuclear point sources.
Molecular hydrogen emission elongated about PA $\sim 100^\circ$ is detected 
in both PSF subtracted and non subtracted line emission maps.
The most straightforward interpretation
for the extended H$_2$ emission is that it lies in a 100 pc diameter `disk'
with a major axis PA $\sim 100^\circ$.  This `disk'
is not aligned with the ionization cone observed
in [OIII] at PA $\sim 30^\circ$ or that
of the 18cm radio emission at PA $\sim -10^\circ$ (\cite{mundell}).
The position angle of the H$_2$ emission 
also differs from that of the major axis of the galaxy (PA $\sim 155^\circ$).
There is evidence from the CO velocity field
that a warped molecular disk is present in this region (\cite{schinnerer}).
The color map shown in Fig.~1 suggests that extinction is
large along the axis of the molecular hydrogen emission.

\subsubsection{NGC~2992}
H$_2$ emission is extended
along a PA $\sim 20^\circ$ similar to that of the major
axis of the galaxy.  The most straightforward interpretation
is that it lies in a 800 pc diameter disk in the plane of the galaxy.
H$\alpha$+[NII] emission is primarily detected above and
below the plane of the galaxy (\cite{wehrle}).

\subsubsection{NGC~5643}

Strong H$_2$ emission is observed near
the nucleus as well as extended emission coincident
with some of the features seen in line emission such
as [OIII].  The overall shape of the line emission
is similar to but not coincident with
dust lanes seen in the optical/near-infrared (F814W/F160W)
color map (\cite{quillen99}).
East of the nucleus the radio jet (PA $= 87^\circ$) 
lies along the southern edge of the ionization cone (\cite{simpson}).

\subsubsection{NGC~2110}

The nucleus appears to be quite red in our visible/near-infrared
color map (see Fig.~4) and so a straight forward subtraction
of the F200N image from the F196N image results in over subtraction
of the nucleus.  In the 1-0~S(3) line emission map we
detect extended features near the nucleus at about $1''$
from the nucleus and in a large loop about $3''$ from the nucleus 
similar in position and shape to that observed in the 
H$\alpha$+[NII] and [OIII]
line emission maps (which are discussed in \cite{mulchaey94a}).  
The overall shape of the line emission
is similar to but not coincident with
features seen in the optical/near-infrared (F814W/F160W)
color map (\cite{quillen99}) and avoids regions of 
emission from the S-shaped radio jet at 
PA $\sim 10^\circ$ (\cite{mulchaey94a}).

\subsubsection{MKN~1066}
MKN~1066 has narrow line emission (e.g., [OIII])
in two narrow features that are aligned with radio emission
from linear jets at PA $=134^\circ$ (\cite{bower95}).
Molecular hydrogen emission is detected in the same region
as the brightest [OIII] emission but appears to be somewhat 
more extended in the direction perpendicular to the radio jets.

\subsubsection{NGC~4945}
The lack of [OIII] emission and low 3$\micron$ (L band) flux density
suggest that a starbursting disk (rather than a low luminosity AGN)
is responsible for the conical
shaped cavity observed in the visible and near-IR images
of NGC~4945 (\cite{moorwood}).
The cavity is interpreted to be a region where
a bipolar superwind or bubble has evacuated denser
material.   Our H$_2$ image resembles 
that \cite{moorwood_}, though more structure is seen.
The morphology suggests that it forms a boundary
between denser gas and a 100 pc scale `bubble' rather than 
a cone.  Extinction on the southern side of the galaxy
is so high at $2 \micron$ that we would not expect
to see molecular hydrogen emission from a southern bubble or cone
if it exists.

\section{Extended emission}

We observed extended H$_2$ emission from
the 1-0 S(1) or S(3) line in 6 Seyfert galaxies: 
NGC~5643, NGC~3227, NGC~2992,
NGC~2110, MKN~1066 and  NGC~4945.   
We do not resolve extended emission in the other galaxies.
In MKN~938 this might be because of its distance.
For MKN~6, NGC~5506 and NGC~5506, the 
nuclear sources are so bright that
it is difficult to observe any extended structure.
Future attempts at modeling the
point spread function might yield detections of molecular
hydrogen emission, particularly in MKN~6 which has
extended [OIII] on a scale that we could resolve
if the nuclear source could be subtracted reliably.

\subsection{H$_2$ emission in a dense `disk'}

In both NGC~3227 and  NGC~2992 molecular hydrogen emission
is coincident with extremely red continuum colors
suggesting that it is associated with large quantities of
dense molecular material.  The morphology of the H$_2$ emission
in both cases is most easily interpreted as emission
from a `disk' of 100 and 800 pc diameter in NGC 3227 
and in NGC 2992 respectively.  In both galaxies line emission
from ionized gas and 6cm radio emission (\cite{wehrle}, \cite{mundell})
are observed along a different axis than the H$_2$ emission.
Because this emission 
is associated with high levels of extinction, and not clearly
associated with either the ionization cones or radio jets,
it is unlikely that 
shocks or ionization associated with the jets and ionization cones
can directly cause the excitation. 

In NGC~3227 one possible interpretation is that the molecular
hydrogen emission originates from the edge of a circumnuclear
disk which has been observed in carbon
monoxide submillimeter interferometry (\cite{schinnerer}).
However in NGC~3327, variations in the distribution and velocity components
observed in [OIII] and H$\alpha$ + [NII] suggest
that star formation is vigorous in its central regions (\cite{gonzalez})
and so that processes associated with star formation
(hot stars and supernovae, e.g.,~\cite{davies3})
could also provide the H$_2$ excitation.

NGC~2992 also has evidence for active star formation. 
The galaxy has large scale outflows observed in H$\alpha$
and soft X-rays which either could be driven by jets or by a starburst
(\cite{colbert}).  H$\alpha$ emission is observed hundreds of pcs from
the nucleus of the galaxy possibly along spiral
arms (e.g., \cite{wehrle}).  
The extent and narrow scale height of the H$_2$ emission 
makes it similar to that observed in the prototypical starburst 
galaxy M82 (\cite{m82}).

\subsection{The H$_2$ to H$\alpha$ ratio}

For NGC~2110, NGC~5643, and MKN 1066,  
$H_2$ emission is associated with
features observed in H$\alpha$+[NII] line emission maps.
We can use the strength of the observed H$\alpha$ emission
to estimate the local radiation flux of ionizing photons.  
If the H$_2$ emission results from UV fluorescence then this flux 
is directly related (via a spectral index in the UV) 
to the flux of UV photons that can disassociate
molecular hydrogen (in the range 912-1100\AA).  Models
(e.g., \cite{vandish}; \cite{puxley}) 
predict 1-0~S(1) molecular hydrogen line strengths
as a function of the incident UV radiation field and the density
of the molecular clouds.  This results in an H$\alpha$ to 1-0~S(1)
line flux ratio typical of HII regions  
(or as used in \cite{puxley} a Br$\gamma$ to 1-0~S(1) ratio).
Here we follow the approach outlined in \cite{davies_} to determine
if UV fluorescence is a feasible H$_2$ excitation mechanism.

Using the surface brightness in the H$\alpha$ line 
($\sim 10^{-14}$ erg s$^{-1}$ cm$^{-2}$ arcsec$^{-2}$  
along the northern loop $3''$ from the nucleus
from \cite{mulchaey94a} in NGC~2110, 
$\sim 5 \times 10^{-15}$ erg s$^{-1}$ cm$^{-2}$ arcsec$^{-2}$ 
in the diffuse emission
along the dust lane from \cite{simpson} in NGC~5643
and $10^{-13}$ erg s$^{-1}$ cm$^{-2}$ arcsec$^{-2}$
on the north-west side of MKN 1066 from \cite{bower95}),
setting the Balmer recombination rate equal to the 
number of ionizing photons (as in \cite{mulchaey94a}) 
and using a UV spectral index of 1.6 (as in \cite{davies}) 
we estimate a UV radiation flux between 912 and 1100\AA
of 50, 22, and 400 times (for NGC~2110, NGC~5643 and MKN~1066 respectively) 
the solar neighborhood value of $6 \times 10^{-4}$ erg s$^{-1}$ cm$^{-2}$.
From this and using the models of \cite{vandish_}
we would predict 1-0~S(1) line fluxes of 
$7 \times 10^{-17}$ for NGC~2110,   
$5 \times 10^{-17}$ for NGC~5643 and 
$2 \times 10^{-16}$ erg s$^{-1}$ cm$^{-2}$ arcsec$^{-2}$ for MKN~1066 
(assuming $n_H = 10^4 {\rm cm}^{-3}$).
This corresponds to the ratio of the 1-0~S(1) to H$\alpha$ flux density
ratio of $\sim 10^{-2}$.  This is equivalent
to a 1-0 S(1)/Br$\gamma$ flux ratio $\sim 1$ and agrees with
the higher density models of \cite{puxley_}.
From our H$_2$ emission maps 
we estimate the surface brightness to be  
$10, 6$ and $30$ times higher than that predicted above 
for UV fluorescence in NGC~2110, NGC~5643 and MKN~1066 respectively.
While extinction of H$\alpha$ may to some extent reconcile the flux estimated
for UV fluorescent excitation with that observed we suspect that 
UV fluorescence cannot be the dominant H$_2$ excitation process in any of
these regions.
H$_2$ excitation by slow shocks is then a more likely process 
(e.g., \cite{koornneef}; \cite{veilleux}; \cite{ruiz}).



\subsection{H$_2$ Emission in the extended narrow line region}

In NGC~2110, NGC~5643 and MKN~1066 we detect emission from 
molecular hydrogen which appears to be associated and
coincident with features observed
in [OIII] and H$\alpha$ + [NII] line emission maps tracing ionized gas.  
The line widths H$_2$, while smaller than the widths of [FeII] 
(\cite{ruiz}; \cite{veilleux})
are still significantly higher than that expected for the low velocity
shocks capable of heating the molecular hydrogen without
disassociating the molecules ($\lesssim 50$ km/s). 
If the H$_2$ emission results from slow shocks then 
these shocks may be produced by collisions in a
multiphase interstellar medium or in other words a medium
with patches of molecular material 
embedded in a more diffuse atomic medium (e.g., \cite{heckman86}).
The association of emission from ionized
gas with emission from molecular hydrogen 
implies that a multiphase medium must exist in or near
these cones.  

In NGC~5643 and NGC~2110 the line emission is
also near dust lanes observed in optical/near-infrared color maps
(\cite{quillen99}).
We see that H$_2$ emission is coincident with H$\alpha$ + [NII] and [OIII]
emission,  however the dust lanes seen in the color maps, though
similar in shape, appear to be offset from the line emission.
If the morphology and orientation of the cone is influenced
by the density distribution of the ambient media
(e.g., as simulated by \cite{mulchaey96}), then we would expect that 
the line emission would peak at a position on the sky that is
somewhat different than the maximum extinction or gas density.
In fact the proximity of dust lanes with the line emission
is good evidence for a dependence of the line excitation
on the density of the ambient medium.
We might hope that any future set of models for the excitation processes
occurring in the ionization cone might be constrained by 
the observed association and offsets.

Since the UV radiation field implied by the H$\alpha$+[NII] and [OIII]
emission is relatively intense we expect that molecular gas
should be disassociated relatively quickly.  Either the molecular
material exists outside the UV radiation field, is self-shielded
or it must be replenished on timescales of $\sim 10^6 -10^7$ years.
The proximity of molecular material also suggests that
that dust is present in the ionization cone.
This possibility could be tested with future high resolution
mid-IR imaging.

\section{Correlations between H$_2$ emission and 
radio and hard X-ray flux}

By comparing the H$_2$ flux with that in hard X-rays and
6cm radio,
previous spectroscopic studies (\cite{veilleux}, \cite{koornneef}, 
\cite{vanzi3} and \cite{ruiz}) have concluded that more than
one emission mechanism is likely to cause the molecular hydrogen
emission in Seyfert galaxies.  Our imaging study has shown
that this is indeed likely.  However these
studies were hampered by small samples
so we take the opportunity here to 
combine the samples of these previous studies.
Molecular hydrogen measurements from \cite{ruiz_} are listed in Table 3.

In Fig.~11 we plot 6cm and hard X-ray fluxes vs molecular hydrogen
fluxes.  A weak correlation noted previously (\cite{ruiz},
\cite{forbes}, \cite{koornneef}) between 6cm
and molecular hydrogen emission is seen.  We note
that the scatter is much larger
than that observed in [FeII] vs 6cm flux (e.g., 
\cite{forbes}; \cite{simpsonfe2}).  For most galaxies
the hard X-ray flux can heat the molecular gas
sufficiently to result in the observed molecular hydrogen emission,
however the lack of correlation between the hard X-ray and molecular hydrogen
rules out this mechanism as the dominant one.

The weak correlation between H$_2$ and 6cm emission and lack
of correlation with the hard X-ray flux suggests 
that more than one process causes the molecular hydrogen emission
in Seyfert galaxies.  Our imaging study suggests that
in some of these galaxies
a substantial fraction of the emission could be associated
with star formation.  Unfortunately in starbursting galaxies 
the flux of H$_2$ emission does not correlate
with any other indicator suggesting
that processes other than supernovae
and young stars are impoartant (C.~Engelbracht, private communication;
\cite{vanzi}).
To test this we looked for (and failed to find) any correlation between 
the IRAS 12-25$\micron$ spectral index (which should
be dependent on fraction of infrared emission from the nucleus)
and the ratio of H$_2$ emission and hard X-ray flux 
and the ratio of H$_2$ and 6cm fluxes.

\section{Summary and Discussion}

In this paper we have presented NICMOS/HST imaging of molecular
H$_2$ in the 1-0 S(1) or 1-0 S(3) emission in Seyfert galaxies
detecting extended emission in 6/10 galaxies.  
We did not detect extended emission in MKN~938 possibly
because of its distance and NGC~5506, IC~4329A and MKN~6
probably because of the luminosity of their nuclear
sources.   Improvements in techniques for subtraction of
NICMOS point spread function (e.g., \cite{krist}) may make it possible
to detect extended molecular hydrogen emission from these images,
particularly in MKN~6.

In NGC~2992 and NGC~3227 the molecular hydrogen emission
is from 800 and 100 pc diameter `disks' (respectively) which
are not coincident with [OIII] or H$\alpha$ emission and are
near high levels of extinction.
In NGC~4945 the molecular hydrogen emission appears
to be from the edge of a 100 pc superbubble.
In these 3 galaxies the molecular gas
could be excited by processes associated with local star formation.
In NGC 3227 the emission might arise from
the edge of a circumnuclear torus or warped disk.

In NGC~5643, NGC~2110 and MKN~1066, molecular hydrogen emission
is detected in the extended narrow line region on scales
of a few hundred pc from the nucleus.  Emission is
found in the ionization cones coincident with [OIII] and H$\alpha$+[NII] line
emission.  This emission is also
near dust lanes observed in the visible to near-infrared
color maps.  The coincidence of H$_2$ emission with that
from ionized gas implies that a multiphase medium
is near or within the ionization cone.
The proximity to dust lanes seen in the color maps
suggests that processes causing
the line emission are dependent on the density of the ambient
media.  The high 1-0~S(1) or S(3) H$_2$ to H$\alpha$
flux ratio suggests that shock excitation  (not UV fluorescence)
of molecular hydrogen is
the dominant excitation process in these extended features.

We have compiled the spectroscopic observations of  molecular hydrogen
(from \cite{veilleux}, \cite{koornneef} and \cite{ruiz})
to look for correlations between 6cm radio and hard X-ray flux.
We find no correlation between H$_2$ emission and hard X-ray flux
confirming the results of these previous studies, 
and ruling out heating by X-rays
as the dominant source of H$_2$ excitation.
There is only a weak correlation between H$_2$ emission flux and
6cm flux density.
Both our imaging study and this compilation
confirm the finding of previous spectroscopic studies (\cite{ruiz},
\cite{koornneef}, \cite{veilleux}, \cite{forbes}):
no single mechanism is likely to be responsible
for the molecular hydrogen excitation in Seyfert galaxies.
Future higher resolution observations may allow us to separate
between likely processes and isolate molecular hydrogen 
emission from the circumnuclear environment.

\acknowledgments

Support for this work was provided by NASA through grant number
GO-07869.01-96A
from the Space Telescope Institute, which is operated by the Association
of Universities for Research in Astronomy, Incorporated, under NASA
contract NAS5-26555.
We also acknowledge support from NASA project NAG-53359.
We acknowledge helpful discussions and correspondence with  
A.~Eckart, E.~Schinnerer, C.~Kulesa, G.~Bower and A.~Cotera.


\clearpage


\vfill\eject

\begin{figure*}
\caption[junk]{
a) NICMOS Camera 2 F160W ($1.60\mu$m)image of NGC~3227.
Contours 0.25 mag apart are overlayed with the faintest
white counter roughly equivalent to 14.0 mag/''$^2$ in Johnson H band.
b) Visible/near-IR color map constructed from F606W and F160W images.
Colors range from V-H$=2.6$ (approximately Johnson bands; white) 
to  V-H$=4.1$ (black).
c) F606W ($0.606\micron$) WFPC2 image. Contours are 0.25 mag apart.
Note that the nucleus is saturated in this image.
d) 1-0~S(1) H$_2$ line emission.  This image was constructed
using PSF subtracted images.  Dark colors refer to brighter emission.
$1''$ is equivalent to 75 pc assuming $H_0 = 75$km s$^{-1}$ Mpc$^{-1}$.
North is up and east is to the left.
}
\end{figure*}

\begin{figure*}
\caption[junk]{
a) NICMOS Camera 2 F205W ($2.05\mu$m) image of NGC~2992.
Contours 0.25 mag apart are overlayed with the faintest
white counter roughly equivalent to 4.61mJy/$''^2$.
b) Visible/near-IR color map constructed from F606W and F205W images.
Colors range from V-K$=4$ (approximately Johnson bands; white)
to  V-K$= 7$ (black).
c) F606W ($0.606 \micron$) WFPC2 image.  Contours are 0.25 mag apart.
Note that the nucleus is saturated in this image.
d) 1-0~S(3) H$_2$ line emission constructed from NICMOS Camera 3 narrow
band images.  Dark colors refer to brighter emission.
$1''$ is equivalent to 153 pc (for $H_0 = 75$km s$^{-1}$ Mpc$^{-1}$).
}
\end{figure*}

\begin{figure*}
\caption[junk]{
a) NICMOS Camera 2 F160W ($1.60\mu$m) image of NGC~5643.
Contours 0.25 magnitude apart are overlayed.  The faintest
white contour corresponds to 14.0 mag/$''^2$ in Johnson H band.
b) Visible/near-IR color map constructed from F814W and F160W images.
Colors range from I-H$=1.3$ (approximately Johnson bands; white)
to  I-H$=2.3$ (black).
c) H$\alpha$ + [NII] emission (from WFPC2 images; see Simpson et al.~1997).
d) 1-0~S(3) H$_2$ line emission map constructed from NICMOS 
Camera 3 narrow band images. 
Dark colors refer to brighter emission.
$1''$ is equivalent to 78 pc (for $H_0 = 75$km s$^{-1}$ Mpc$^{-1}$).
}
\end{figure*}

\begin{figure*}
\caption[junk]{
a) NICMOS Camera 3 F200N ($2.00\mu$m) image of NGC~2110.
Contours 0.25 magnitude apart are overlayed with the faintest
white counter roughly equivalent to 1.9mJy/$''^2$. 
b) Visible/near-IR color map constructed from F606W and F200N images.
Colors range from V-K$=4$ (approximately Johnson bands; white)
to  V-K$=5.4$ (black).
c) H$\alpha$ + [NII] emission (from PC images; see Mulchaey et al.~1994).
d) 1-0~S(3) H$_2$ line emission constructed from NICMOS
Camera 3 narrow band images.  Dark colors refer to brighter emission.
$1''$ is equivalent to 148 pc (for $H_0 = 75$km s$^{-1}$ Mpc$^{-1}$).
}
\end{figure*}

\begin{figure*}
\caption[junk]{
a) NICMOS Camera 2 F160W ($1.60\mu$m) image of MKN~1066.
Contours 0.5 magnitude apart are overlayed with the faintest
white contour roughly equivalent to 14.0 mag/$''^2$ in Johnson
H band. 
b) Visible/near-IR color map constructed from F606W and F160W images.
Colors range from V-H$=2.5$ (approximately Johnson bands; white)
to  V-H$=4.5$ (black).
Note that the F606W image is contaminated with emission from 
H$\alpha$.  Light corresponds to blue
colors or regions of bright H$\alpha$ emission and
dark corresponds to red colors.
c) [OIII] line emission (from PC images; see Bower et al.~1995).
d) 1-0~S(1) H$_2$ line emission constructed from NICMOS
Camera 2 narrow band images.
Dark colors refer to brighter emission.
$1''$ is equivalent to 238 pc (for $H_0 = 75$km s$^{-1}$ Mpc$^{-1}$).
}
\end{figure*}

\begin{figure*}
\caption[junk]{
a) NICMOS Camera 2 F205W ($2.05\mu$m) image of NGC~4945.
Contours 0.25 mag apart are overlayed with the faintest
white counter roughly equivalent to 4.61mJy/$''^2$. 
b) 1-0~S(1) H$_2$ line emission constructed from NICMOS
Camera 2 narrow band images.
Dark colors refer to brighter emission.
$1''$ is equivalent to 36.3 pc (for $H_0 = 75$km s$^{-1}$ Mpc$^{-1}$).
}
\end{figure*}

\begin{figure*}
\caption[junk]{
a) NICMOS Camera 2 F205W ($2.05\mu$m image) of MKN~938.
Contours 0.25 magnitude apart are overlayed with the  faintest
white contour corresponding to 4.21 mJy/$''^2$.
b) Visible/near-IR color map constructed from F606W and F205W images.
Colors range from V-K$=2.8$ (approximately Johnson bands; white)
to  V-K$= 4.8$ (black).
c) F606W ($0.606\micron$) WFPC2 image.
d) 1-0~S(1) H$_2$ line emission constructed from NICMOS
Camera 2 narrow band images.
We do not detect any extended emission in molecular hydrogen.
Dark colors would refer to brighter emission.
$1''$ is equivalent to 384 pc (for $H_0 = 75$km s$^{-1}$ Mpc$^{-1}$).
}
\end{figure*}

\begin{figure*}
\caption[junk]{
a) NICMOS Camera 2 F160W ($1.60\micron$) image of NGC 5506.
b) Continuum subtracted narrow band image of the
1-0~S(3) H$_2$ line constructed from NICMOS
Camera 3 narrow band images.  The images
are dominated by the Seyfert 1 point source and we
detect no extended line emission.
$1''$ is equivalent to 120 pc (for $H_0 = 75$km s$^{-1}$ Mpc$^{-1}$).
}
\end{figure*}

\begin{figure*}
\caption[junk]{
a) NICMOS Camera 2 narrow band continuum F212N ($2.12\micron$) image 
of IC~4329A.
b) Continuum subtracted narrow band image of the 1-0~S(1) H$_2$ line.  
The images are dominated by the Seyfert 1 point source
and we detect no extended line emission.
$1''$ is equivalent to 311 pc (for $H_0 = 75$km s$^{-1}$ Mpc$^{-1}$).
}
\end{figure*}

\begin{figure*}
\caption[junk]{
a) NICMOS Camera 1 F160W ($1.60\micron$) image of MKN 6.
b) F606W ($0.606\micron$) WFPC2 image.   We have displayed
this image to show the dust lane to the north.  The nucleus
is saturated in this image.
c) [OIII] line emission (from FOC images; see Capetti et al.~1995).
d) Continuum subtracted narrow band image of the 1-0~S(1) H$_2$ line.
The images are dominated by the Seyfert 1 point source
and we detect no extended line emission.
$1''$ is equivalent to 358 pc (for $H_0 = 75$km s$^{-1}$ Mpc$^{-1}$).
}
\end{figure*}

\begin{figure*}
\caption[junk]{
a)  Molecular hydrogen line flux vs 6 cm radio emission.
Seyfert 1s are shown as solid triangles and Seyfert 2s as open
triangles.   We see only a week correlation between 6cm and molecular hydrogen
flux. The weakness of the correlation is similar
to that presented in Veilleux et al.~(1997) and Ruiz (1997).
We note that a correlation was also reported by Forbes \& Ward (1993) and
Koornneef \& Israel (1996).
H$_2$ and 6cm fluxes are compiled from Koornneef \& Israel (1996), Ruiz (1997),
Veilleux et al.~(1997), and  Moorwood \& Oliva (1988).
b)  Molecular hydrogen line flux vs hard X-ray flux.
Points below and to the right of the solid line have
sufficient X-ray flux to account for
the heating of the molecular gas (as estimated by Lepp \& McCray 1983,
$F(HX) = 400 f(H_2 \lambda 2.121$); this plot is similar to
that shown in Veilleux et al.~1997).
However the lack of correlation between molecular hydrogen emission
and X-ray flux suggests that
heating by X-rays is unlikely to be the dominant heat source
for the molecular hydrogen emission.
For the Seyfert 2s hard X-ray fluxes are compiled from Bassani et al.~(1999)
and for the Seyfert 1s from Reynolds (1997).
}
\end{figure*}

\clearpage

%


\begin{deluxetable}{lclclcccccccc}
\footnotesize
\tablecaption{Sample}
\tablehead{
\colhead{Galaxy}                           &
\colhead{Inclination}                      &
\colhead{Type}                             &    
\colhead{Flux H$_2$}                       &          
\colhead{$z$}                              &          
\colhead{Obs. wave}                        &          
\colhead{Camera}                           &
\colhead{Line}                             &
\colhead{Continuum}                          \cr
\multicolumn{1}{l}{(1)} &
\colhead{(2)}                              &
\colhead{(3)}                              &
\colhead{(4)}                              &
\colhead{(5)}                              &
\colhead{(6)}                              &
\colhead{(7)}                              &
\colhead{(8)}                              &
\colhead{(9)}                              
} 
\startdata
%
 NGC 3227   & 45 & Sy1.5& 29.0 & 0.00386 & 2.129 & NIC2 & F212N & F215N \cr
 NGC 2992   & 70 & Sy2  & 12   & 0.0079  & 1.973 & NIC3 & F196N & F200N \cr
 NGC 2110   & 40 & Sy2  & 9.25 & 0.00762 & 1.972 & NIC3 & F196N & F200N \cr
 NGC 5643   & 30 & Sy2  & 21   & 0.00400 & 1.965 & NIC3 & F196N & F200N \cr
 NGC 4945   & 80 & Sy2  & 129  & 0.00187 & 2.122 & NIC2 & F212N & F215N \cr
 Mkn 1066   & 55 & Sy2  & 13.6 & 0.0123  & 2.147 & NIC2 & F215N & F212N \cr
 Mkn 938    & 70 & Sy2  & 17.3 & 0.01978 & 2.163 & NIC2 & F216N & F212N \cr
 Mkn 6      & 50 & Sy1.5& 19.5 & 0.01847 & 2.161 & NIC2 & F216N & F212N \cr
 IC 4329A   & 75 & Sy 1 & 11.8 & 0.01605 & 2.155 & NIC2 & F215N & F212N \cr
 NGC 5506   & 70 & Sy1.9& 10.6 & 0.00618 & 1.970 & NIC3 & F196N & F200N \cr
\enddata
\tablenotetext{}{
(1) Galaxy;
(2) Galaxy inclination in degrees where $0 =$ face-on;
(3) Seyfert type;
(4) 1-0 S(1) 2.1213$\micron$ line flux in units of
$10^{-15}$ erg cm$^{-2}$ s$^{-1}$
compiled in Ruiz (1997), Veilleux et al.~(1997), or Koornneef \& Israel
(1996).
All but NGC 5643, Mkn1066, NGC 4945 are from Ruiz (1997) and
fluxes were extracted from a $1.2''$ slit.
NGC 4945 is from Koornneef \& Israel (1996) and was from a $6$x$6''$ aperture.
MKN 1066 flux is from Veilleux et al.~(1997) and was from a $1.5$x$1.5''$ 
aperture.
(5) Redshift;
(6) Observable wavelength (redshifted) for the 1-0 S(1) 2.1213$\micron$
or 1-0 S(3) 1.9576$\micron$ line;
(7) NICMOS camera used for narrow band imaging;
(8) Narrow band filter containing the H$_2$ line.
(9) Continuum filter.
}
\end{deluxetable}


\begin{deluxetable}{lllllcccccccc}
\footnotesize
\tablewidth 4.5truein
\tablecaption{1-0 S(1) and 1-0 S(3) H$_2$ fluxes}
\tablehead{
\colhead{Galaxy} &
\multicolumn{2}{c}{Region} &
\colhead{Flux}  &
\colhead{Surface brightness} 
}
\startdata
NGC 5643 & eastern feature   & $1.2'' < r < 8.16''$   & 6   & 3 \nl
         & western feature   & $1.2'' < r < 8.16''$   & 2   &  \nl
         & nuclear           & $r<1.2''$              & 34  &  \nl
NGC 2110 & near the nucleus  & $0.6''< r < 1.6''$     & 8   &  \nl
         & northern loop     & $2.0 < r < 4.0''$      & 2   & 6 \nl
NGC 2992 & nuclear           & $  r< 0.6''  $         & 33  & \nl
         & disk              & $0.6''< r < 6.5''$     & 68  & \nl
NGC 3227 & disk              & $0.4 < r< 1.3''$       & 10  & \nl
MKN 1066 & total             & $r<1.1''$              & 6   & \nl
         & western feature   & $r\sim 0.5''$          &     & 70 \nl
NGC 4945 & total             &                        & 170 & \nl 
\enddata
\tablenotetext{}{
NGC~5643, NGC~2110, and NGC~2992 fluxes are for the 1-0 S(3) line.
For the remaining galaxies we list fluxes in the 1-0 S(1) line.
Fluxes are listed in units of $10^{-15}$ erg cm$^{-2}$ s$^{-1}$
and surface brightnesses in units of 
$10^{-16}$ erg cm$^{-2}$ s$^{-1}$$''^{-2}$.
}
\end{deluxetable}


\begin{deluxetable}{lcc}
\footnotesize
\tablewidth 3.0truein
\tablecaption{1-0S(1)H$_2$ fluxes 
and FWHM  from Ruiz (1997).}
\tablehead{
\colhead{Galaxy} &
\colhead{flux} &  
\colhead{FWMH} \nl
\colhead{} &
\colhead{$10^{-14}$ erg cm$^{-2}$ s$^{-1}$} &
\colhead{km s$^{-1}$} 
}
\startdata
Mkn~334   & $ 0.91$   & 130    \nl
Mkn~335   & $ 0.13$   & 102    \nl
Mkn~938   & $ 1.73$   & 537    \nl
Mkn~348   & $<0.04$   &\nodata \nl
Mkn~1040  & $ 0.24$   & 185    \nl
NGC~1275  & $ 3.71$   & 382    \nl
Mkn~1095  & $ 0.51$   & 381    \nl
NGC~2110  & $ 0.93$   & 447    \nl
MCG-8-11-11 &$1.19$   & 386    \nl
NGC~2273  & $ 0.68$   & 251    \nl
Mkn~6     & $ 1.95$   & 234    \nl
Mkn~376   & $ <0.02$  &\nodata \nl
Mkn~79    & $ <0.05$  &\nodata \nl
Mkn~704   & $ <0.06$  &\nodata \nl
NGC~2992  & $ 0.41$   & 327    \nl
NGC~3227  & $ 2.90$   & 331    \nl
NGC~3516  & $<0.01$   &\nodata \nl
NGC~4051  & $ 0.87$   & 253    \nl
NGC~4151  & $ 0.20$   & 232    \nl
Mkn~766   & $ 0.24$   & 214    \nl
IC~4329A  & $ 1.18$\tablenotemark{a}   & 332 \nl
NGC~5506  & $ 1.06$   & 315    \nl
Mkn~817   & $<0.02$   &\nodata \nl
NGC~7469  & $ 0.86$   & 309    \nl
Mkn~533   & $ 0.48$   & 268    \nl
\enddata
\tablenotetext{a}{Derived from Giannuzzo et al.~(1995).}
\end{deluxetable}

\end{document}